\documentclass[11pt]{article}
\usepackage{amsmath,amssymb,amsfonts,epsf,graphicx}
\usepackage[nosort]{cite}
\usepackage[margin=1in]{geometry}

\makeatletter \@addtoreset{equation}{section} \makeatother

\newcommand{\fft}[2]{{\frac{#1}{#2}}}

\def\nn{\nonumber}

\newcommand{\be}{\begin{equation}}
\newcommand{\ee}{\end{equation}}
\def\ba{\begin{array}}
\def\ea{\end{array}}

\def\fft#1#2{\frac{#1}{#2}}

\def\sst#1{{\scriptscriptstyle #1}}

\def\dalemb#1#2{{\vbox{\hrule height .#2pt
        \hbox{\vrule width.#2pt height#1pt \kern#1pt
                \vrule width.#2pt}
        \hrule height.#2pt}}}

\newcommand{\bea}{\begin{eqnarray}}
\newcommand{\eea}{\end{eqnarray}}

\def\0{{\sst{(0)}}}
\def\1{{\sst{(1)}}}
\def\2{{\sst{(2)}}}
\def\3{{\sst{(3)}}}
\def\4{{\sst{(4)}}}
\def\5{{\sst{(5)}}}
\def\6{{\sst{(6)}}}
\def\7{{\sst{(7)}}}
\def\8{{\sst{(8)}}}

\begin{document}

\begin{center}\ \\ \vspace{60pt}
{\Large {\bf Strongly-Coupled Quarks and Colorful Black Holes in AdS/CFT}}\\ 
\vspace{30pt}

MD Razikul Islam$^{1}$ and Justin F. V\'azquez-Poritz$^{2}$
\vspace{20pt}

$^1${\it Physics Department\\ New York City College of Technology, The City University of New York\\ 300 Jay Street, Brooklyn NY 11201, USA.}

\vspace{10pt}
$^2${\it The Graduate School and University Center, The City University of New York\\ 365 Fifth Avenue, New York NY 10016, USA}\\

\vspace{20pt}

{\tt jvazquez-poritz@citytech.cuny.edu}\\
{\tt md.islam3@mail.citytech.cuny.edu}

\end{center}

\vspace{30pt}

\centerline{\bf Abstract}

\noindent Small black holes potentially created by the Large Hadron Collider could typically carry color charges inherited from their parton progenitors. The dynamics of quarks near such a black hole depends on the curved spacetime geometry as well as the strong interaction due to the color charge of the black hole. We use the AdS/CFT Correspondence to study the behavior of strongly-coupled quarks near a color-charged black hole. The supergravity background consists of a six-dimensional Schwarzschild-black string AdS soliton, for which the bulk horizon extends from the AdS boundary down to an infra-red floor. By going to higher energy scales, the regime of validity of the classical supergravity background can be extended closer to the singularity than might be expected from the four-dimensional perspective. We study the resulting behavior of quarks and compute the rate at which a quark rotating around the black hole loses energy as a function of its mass parameter and angular velocity. We also analyze the behavior of quark-antiquark string configurations and discuss features such as the quark-antiquark screening length.

\thispagestyle{empty}

\pagebreak



\section{Introduction}

Large extra dimensions \cite{large1,large2} could potentially reduce the scale of strong quantum gravity to TeV order. Besides resolving the hierarchy problem, this opens the possibility that small black holes could be produced by the Large Hadron Collider \cite{argyres}, and may form within strange stars, which consist entirely of deconfined quark matter \cite{strange}. It is unlikely that such quantum black holes can be described in terms of classical gravity  \cite{quantum}; even the notion of an event horizon is suspect due to potentially large curvature in the vicinity of the would-be horizon. Such a quantum black hole could carry electric charge, color charge and angular momentum inherited from the partons that created it. BlackMax is a black hole event generator that can account for all gauge charges and is being used to put constraints on mini black hole production at the LHC \cite{blackmax1,blackmax2}.

The black hole would interact with nearby quarks via gravity as well as color and electrical interactions. However, there are two potential roadblocks for describing such a system. Firstly, the semi-classical approach to gravity is unreliable when the quarks are close to the black hole. Secondly, the color interaction might take place at  strong coupling, in which case conventional quantum field theoretic methods are ill-equipped to describe dynamical processes. We will demonstrate that the AdS/CFT correspondence \cite{ads-cft1,ads-cft2,ads-cft3,agmoo} may be used to evade both of the aforementioned difficulties. 

While we consider the color interactions between quarks and the black hole, we will consider a black hole with only a small amount of color charge, so that the effect of the color charge on the background itself is negligible. This enables us to use a supergravity solution for which the field contributions corresponding to the color charge have not been turned on. We are also making the working assumption that the AdS/CFT correspondence can be applied even though there is a singularity extending out to the boundary.

The dynamics of quarks in strongly-coupled plasmas have been studied extensively by considering open strings near a black brane that is extended along the spacetime directions of the field theory; see for example \cite{liu,herzog,gubser,teaney}. In contrast, a strongly-coupled system of quarks near a black hole corresponds to a supergravity background that contains a black string extending along the bulk radial direction, various possibilities of which have been discussed in \cite{hubeny,hubeny2,hubeny3,caldarelli}. As one might expect, for both of these scenarios there are a number of similarities in the behavior of the quarks, such as quark energy loss and quark-antiquark screening length.

This paper is organized as follows. In section 2, we show that a black string exhibits a nakedly singular surface, which motivates us to consider a Schwarzschild-black string AdS soliton. We discuss this supergravity background, as well as how to embed flavor D8-branes on it once it has been lifted to massive type IIA theory. In section 3, we discuss single-quark string configurations on this background. In particular, we compute the rate at which a quark steadily rotating around a color-charged black hole loses energy as a function of its mass parameter and angular velocity. In section 4, we analyze the behavior of quark-antiquark string configurations and discuss features such as the quark-antiquark screening length. We discuss possible future directions in section 5. In the appendix, we present various generalizations of the Schwarzschild-black string AdS soliton.

\section{Some Preliminaries}

\subsection{A preamble: The five-dimensional black string}

In this paper, we will be focusing on the six-dimensional Schwarzschild-black string AdS soliton. In order to motivate the consideration of this background, we will first consider the black string.

The metric for a Poincar\'e patch of AdS$_5$ can be written as
\be
ds_5^2= \fft{r^2}{R^2}\ ds_4^2+\fft{R^2}{r^2}\ dr^2,
\ee
where $ds_4^2$ is the 4-dimensional Minkowski metric. This corresponds to the vacuum of ${\cal N}=4$ super Yang-Mills theory at large $N$ and large 't Hooft coupling. We can attempt to describe a quantum field theory in the curved spacetime background of a black hole by replacing the four-dimensional slice by the Schwarzschild metric
\be\label{metric4}
ds_4^2=-f dt^2+f^{-1} d\rho^2+\rho^2 \left( d\theta^2+\sin^2\theta\ d\phi^2\right),
\ee
where 
\be
f=1-\fft{2M}{\rho}.
\ee
This black string is a solution to five-dimensional Einstein gravity and was considered as a toy model for a brane-world black hole \cite{hawking}. There is a singularity at $\rho=0$ which runs  all along the $r$ direction. This string-like singularity is shielded behind a horizon at $\rho=2M$. 

Since the geometry is asymptotically locally AdS for $\rho>0$, the black string would be a natural candidate for describing a dual field theory in the curved spacetime background of a black hole. However, as discussed in \cite{hawking}, this solution contains a nakedly singular surface at $r=0$. This can be seen by the fact that the Kretschmann scalar
\be
K=\fft{40}{R^4}+\fft{48M^2 R^4}{\rho^6 r^4},
\ee
diverges at $r=0$. 

We can attempt to cover this singular surface by a horizon, by considering the following metric
\be\label{AdSbh}
ds_5^2= \fft{r^2}{R^2} \left( -\left( 1-\fft{r_h^4}{r^4}\right) f(\rho) dt^2+f(\rho)^{-1} d\rho^2+\rho^2 \left( d\theta^2+\sin^2\theta\ d\phi^2\right) \right)+\fft{R^2}{r^2} \left( 1-\fft{r_h^4}{r^4}\right)^{-1} dr^2.
\ee
The Kretschmann scalar is
\be
K=\fft{40}{R^4}+\fft{48M^2 R^4}{\rho^6 r^4}+\fft{72r_h^8}{R^4 r^8}+\fft{32M^2 r_h^8}{r^6(r^4-r_h^4)\rho^3 (\rho-2M)}.
\ee
For $M=0$, (\ref{AdSbh}) is the metric for the AdS black brane, for which the singular surface at $r=0$ is covered by a horizon at $r=r_h$. However, for $M\neq 0$ the situation worsens since now the surfaces at $r=r_h$ as well as $\rho=2M$ are singular. In order to avoid nakedly singular surfaces, we will consider a Schwarzschild soliton string solution.

\subsection{The six-dimensional Schwarzschild-black string AdS soliton}

We consider a solution to six-dimensional Einstein gravity which is closely related to the AdS soliton \cite{horowitz1}, whose metric is
\be\label{metric5}
ds_6^2=\fft{r^2}{R^2}\left( h dy^2+ds_4^2\right)+\fft{R^2}{r^2} h^{-1}\ dr^2,\qquad h=1-\fft{b^5}{r^5},
\ee
where the $y$ direction is circular and corresponds to a large extra dimension. This is related by a double Wick rotation to the planar Schwarzschild-AdS black hole. Although we consider the simplest scenario in which there is a single extra dimension, this is readily generalized to an arbitrary number of flat extra dimensions. Also, evidence for the existence of a class of generalizations of the AdS soliton with the circle $S^1$ replaced by a sphere $S^m$ was presented in \cite{soliton-sphere}. For the AdS soliton, $ds_4^2$ is the Minkowski metric. A Schwarzschild-black string AdS soliton \cite{hubeny} can be constructed by simply replacing $ds_4^2$ with the Schwarzschild metric (\ref{metric4}).

\begin{figure}[h]
\centering
\includegraphics[width=150mm]{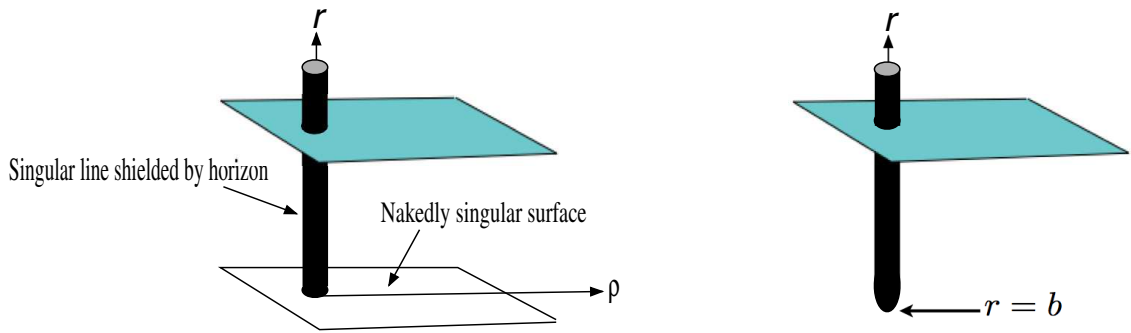}
\caption{The left picture depicts the $b=0$ limit of the Schwarzschild-black AdS soliton, for which there is a nakedly singular surface at $r=0$. The right picture shows that it acquires a smooth cigar-like geometry for $b> 0$. In each case, a probe D-brane is denoted by the turquoise surface.} 
\label{fig7}
\end{figure}

There is a singularity at $\rho=0$ which runs  all along the $r$ direction and is shielded by a horizon at $\rho=2M$. For vanishing $b$ there is a nakedly singular surface at $r=0$, as depicted in the left picture of Figure \ref{fig7}.
This is avoided by restricting the range of the radial coordinate to be $r\ge b>0$. As depicted in the right picture of Figure \ref{fig7}, the $S^1$ smoothly caps off like a cigar geometry at $r=b$, provided that the coordinate $y$ has a periodicity of $\Delta y=4\pi R^2/(5b)$. The Schwarzschild-black string AdS soliton has a geometry that is asymptotically locally AdS$_6$. Thus, as we will discuss in more detail, this is conjectured to describe a strongly-coupled field theory near a black hole, with one extra dimension.

Let us now examine the regime of validity of the classical supergravity background. The Kretschmann scalar is given by
\be
K=\fft{12}{R^4} \left( 5+\fft{20b^{10}}{r^{10}}+\fft{4M^2 R^8}{\rho^6 r^4}\right),
\ee
which is small as long as
\be
R\gg 1,\qquad \rho\gg M \left( \fft{R}{r}\right)^{2/3}.
\ee
Compare this to the perspective of the four-dimensional field theory, for which the black hole background has small curvature for $\rho\gg M$. Thus, at an energy scale $r>R$, the AdS/CFT correspondence enables us to probe regions closer to the black hole than could be done by using the semi-classical approach directly with the four-dimensional theory.

The above six-dimensional background can be embedded within massive type IIA theory \cite{massiveIIA} in ten dimensions by using the Kaluza-Klein reduction given in \cite{6dbh}. The ten-dimensional metric is
\be\label{10dmetric}
ds_{10}^2=s^{-1/3} \left( ds_6^2+g^{-2} (d\alpha^2+c^2 d\Omega_3^2)\right),
\ee
where $s=\sin\alpha$, $c=\cos\alpha$ and the gauge coupling constant $g$ is related to the mass parameter $m$ of the massive Type IIA theory by $m=\sqrt{2}g/3$. Since we will use this as the background geometry in the string action, we have expressed the metric in the string frame. This is a warped product of the six-dimensional space and a hemisphere of $S^4$. The warp factor is singular at the boundary at $\alpha=0$ and the string coupling diverges there since $e^{\Phi}=s^{-5/6}$. 

In the limit in which $M=0$, $b=0$ and the $y$-direction is extended, the above background reduces to a warped product of AdS$_6$ and a hemisphere of $S^4$. This arises in the near-horizon limit of a localized D4-D8 brane configuration \cite{Youm} and has been conjectured in \cite{Ferrara,Brandhuber} to be dual to a five-dimensional ${\cal N}=2$ superconformal theory discussed in \cite{Seiberg,Intriligator}. Turning on the parameter $b$ corresponds to putting in an ``infra-red floor'' for the field theory. The $y$-direction can be made to be compact so that the theory lives on Mink$_4\times S^1$ with supersymmetry breaking boundary conditions imposed along the $S^1$ \cite{horowitz1}. Finally, turning on the parameter $M$ corresponds to this field theory living on the background of a Schwarzschild black hole.

\subsection{Embedding Flavor D8-Branes}
 
The DBI action for a D8-brane is given by
\be
S_{D8}=-T_8\int d^9\sigma\ e^{-\Phi}\sqrt{-\mbox{det}(g_{ab})},
\ee
where $g_{ab}=G_{MN}\partial_a X^M \partial_b X^N$ is the induced metric on the worldvolume of the D8-brane with worldsheet coordinates $\sigma^a$. $G_{MN}$ is the ten-dimensional metric with $\partial_a=\partial /\partial \sigma^a$. We would like to consider the shape of the D8-branes as a function of $r$ on the metric (\ref{10dmetric}) which describes the ten-dimensional lift of the Schwarzschild-black string AdS soliton. 

We can consider an embedding in which the D8-branes extend in four-dimensional spacetime, are wrapped on the internal 4-sphere, and form a curve for slicings of constant $\rho$ described by the function $r(y,\rho,\alpha)$. The DBI action then reads
\be
S_{D8}=-C\int dt d\rho d\alpha dy\ s^{-13/6} c^3 r^5\rho^2 \sqrt{\left( h+\fft{R^4}{r^4h} (\partial_y r)^2\right) \left( 1+\fft{g^2R^2}{r^2 h} (\partial_{\alpha} r)^2\right) \left( 1+\fft{R^4}{r^4 h} (\partial_{\rho} r)^2\right)}, 
\ee
where $C=T_8V_2V_3/(g^4R^5)$, and $V_2$ and $V_3$ are the volumes of the $S^2$ and $S^3$.
The equation of motion is
\bea
&& \partial_{\rho} \left[ s^{-13/6} c^3 r^5 \rho^2 A^{-1} \left( h+\fft{R^4}{r^4h} (\partial_y r)^2\right) \left( 1+\fft{g^2R^2}{r^2h} (\partial_{\alpha} r)^2\right) \fft{R^4f}{r^4h} (\partial_{\rho} r)\right]\nn\\
&+& \partial_{\alpha} \left[ s^{-13/6} c^3 r^5 \rho^2 A^{-1} \left( h+\fft{R^4}{r^4h} (\partial_yr)^2\right) \left( 1+\fft{R^4f}{r^4h} (\partial_{\rho}r)^2\right) \fft{g^2R^2}{r^2h} (\partial_{\alpha}r)\right]\nn\\
&+& \partial_y \left[ s^{-13/6} c^3 r^5 \rho^2 A^{-1} \left( 1+\fft{g^2R^2}{r^2h} (\partial_{\alpha} r)^2\right) \left( 1+\fft{R^4}{r^4h} (\partial_{\rho} r)^2\right) \fft{R^4}{r^4h} (\partial_yr)\right]\nn\\
&-& \partial_r \left[ s^{-13/6} c^3 r^5 \rho^2 A\right]=0,
\eea
where
\be
A= \sqrt{\left( h+\fft{R^4}{r^4h} (\partial_y r)^2\right) \left( 1+\fft{g^2R^2}{r^2 h} (\partial_{\alpha} r)^2\right) \left( 1+\fft{R^4}{r^4 h} (\partial_{\rho} r)^2\right)}.
\ee
For the case $r=r(y)$, the first integral of the equation of motion reduces to
\be
(\partial_yr)^2=\fft{r^4h^2}{r_0^{10} h_0 R^4} \left( r^{10}h-r_0^{10}h_0\right),
\ee
where the D8-branes curve down to the minimal radius of $r=r_0$, and $h_0=h(r_0)$. While there is a ``trivial'' solution with $r_0=0$ which represents disjoint D8-branes, the curved solution is the energetically favored one.

\section{Single-Quark String Configurations}

The dynamics of a classical string are governed by the Nambu-Goto action
\be
S=-\fft{1}{2\pi\alpha^{\prime}} \int d^2\sigma \sqrt{-\mbox{det}(g_{ab})}\,,
\ee
where $g_{ab}=G_{MN}\partial_a X^M \partial_b X^N$ is the induced metric on the string worldsheet. We will consider strings on the background described by the ten-dimensional metric (\ref{10dmetric}), which move purely in the six-dimensional space and lie at a point on the 4-sphere. In fact, it can be shown that the equations of motion require the string to lie at the ``north pole'' of the 4-sphere at $\alpha=-\pi/2$. This is the result of an effective gravitational repulsion on the string that is due to the warp factor. Thus, $X^M$ run over the six coordinates ${t,\rho,\theta,\phi,y,r}$. Since $\alpha^{\prime}$ can be canceled out in the action anyway by using the radial coordinate $u=r/\alpha^{\prime}$, we set $\alpha^{\prime}=1$

\begin{figure}[h]
\centering
\includegraphics[width=150mm]{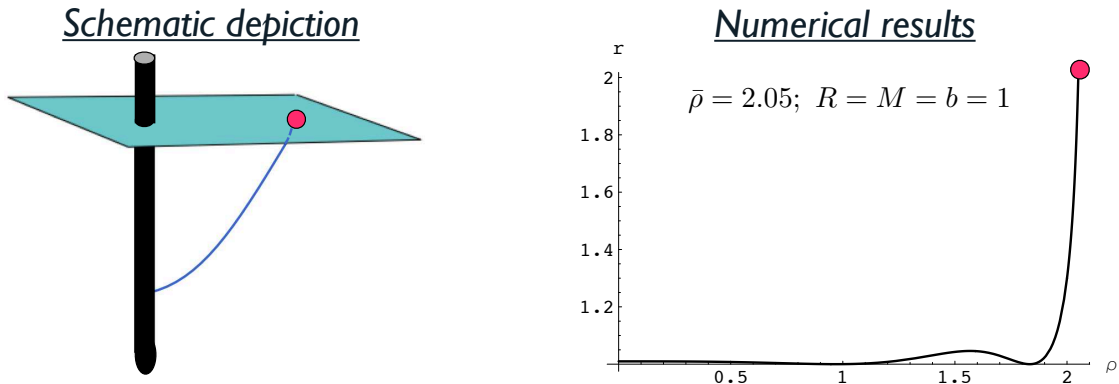}
\caption{The picture on the left is a schematic depiction of the string configuration corresponding to a quark interacting with a color-charged black hole. The corresponding numerical plot is shown on the right.} 
\label{fig2}
\end{figure}

For a single-quark string configuration, an endpoint lies on a probe D-brane and the string goes through the horizon at $\rho=2M$ and approaches the singularity at $\rho=0$, as is schematically depicted in the left picture of Figure \ref{fig2}. Due to the spherical symmetry of the four-dimensional spacetime, a static string configuration lives within the $r-\rho$ plane. However, we will consider a string that is steadily moving in the $\phi$, for which we will allow for the possibility that the string spirals around the $\phi$ direction as well. This string configuration can be described by the worldsheet embedding
\be\label{embedding1}
t=\tau,\qquad \rho=\sigma,\qquad \theta=\fft{\pi}{2},\qquad \phi= \phi(\sigma)+\omega\tau,\qquad r=r(\sigma),
\ee
with the boundary conditions
\be
0\le \tau\le T,\qquad 0\le\sigma\le\bar\rho,\qquad r(\bar\rho)=\bar r,
\ee
where the endpoint of the string is located at $r=\bar r$ and $\rho=\bar \rho$. $\omega$ and $v=\bar\rho\omega$ are respectively the angular velocity and velocity of the string endpoint measured by an observer far from the black hole. On the other hand, the proper velocity $V$ of the string endpoint is given by
$V=v f^{-1/2}|_{\rho=\bar\rho}$. Thus, in order for the string to be timelike, we must have $V<1$ which translates into
\be
v<\sqrt{1-\fft{2M}{\bar\rho}}.
\ee
With the above worldsheet embedding, the action becomes
\be\label{S1}
S=-\fft{T}{2\pi} \int d\rho \sqrt{\left( f-\omega^2\rho^2\right)\left(\fft{r^4}{R^4}f^{-1}+h^{-1}r^{\prime 2}\right)+\fft{r^4}{R^4} f\rho^2\phi^{\prime 2}},
\ee
where $^{\prime}\equiv\partial/\partial\rho$. The equations of motion are given by
\bea\label{eom3}
\fft{d}{d\rho} \left[ A (f-\omega^2\rho^2)h^{-1} r^{\prime}\right] &=& \fft{A}{R^4} \left[ (f-\omega^2\rho^2) \left( 2r^3f^{-1}-\fft{5R^4b^5}{2h^2r^6} r^{\prime 2}\right)+2\rho^2 r^3f\phi^{\prime 2}\right],\nn\\
\fft{d}{d\rho} \left[ Ar^4f\rho^2\phi^{\prime}\right] &=& 0,
\eea
where
\be\label{A}
A^{-1}=\sqrt{(f-\omega^2\rho^2)(r^4f^{-1}+R^4h^{-1}r^{\prime 2})+\rho^2r^4f\phi^{\prime 2}}.
\ee
The second equation in (\ref{eom3}) can be integrated once to give
\be\label{phi}
\phi^{\prime}=\fft{c}{Ar^4f\rho^2},
\ee
where $c$ is an integration constant that parameterizes how much the string spirals around the $\phi$ direction. In particular, $c=0$ corresponds to a string that lives solely in the $r-\rho$ plane. 

The canonical momentum densities associated to the string are given by
\bea
\pi_{\mu}^0 &=& -h_{\mu\nu}\ \fft{(\dot X\cdot X^{\prime})(X^{\nu})^{\prime}-(X^{\prime})^2 (\dot X^{\nu})}{2\pi \sqrt{-G}},\nn\\
\pi_{\mu}^1 &=& -h_{\mu\nu}\ \fft{(\dot X\cdot X^{\prime})(\dot X^{\nu})-(\dot X)^2 (X^{\nu})^{\prime}}{2\pi \sqrt{-G}}.
\eea
The total energy and momentum in the $\phi$ direction of the string are
\be
E=-\int d\sigma\ \pi_t^0,\qquad p_{\phi}=\int d\sigma\ \pi_x^0.
\ee
The string will gain energy through its endpoint at a rate of
\be
\pi_t^1=\fft{\omega c}{2\pi R^2}.
\ee
The string exerts a force on its endpoint, which must be countered by an external force. For a purely radial string, the external force must be radially outwards. If the string spirals around the $\phi$ direction, then we also need to apply a force in the $\phi$ direction given by
\be
\pi_{\phi}^1=\fft{c}{2\pi R^2}.
\ee
From the four-dimensional gravity perspective this is rather surprising, given that the quark is orbiting the black hole at a constant speed and the black hole is not rotating. However, this is a result of the strong interaction between the quark and the black hole.

From (\ref{A}) and (\ref{phi}), we find
\bea
-G &=& \fft{1}{R^4 A^2} = \fft{\rho^2 r^4}{R^4} \left(\fft{f-\omega^2\rho^2}{r^4f\rho^2-c^2}\right) (r^4+R^4fh^{-1}r^{\prime 2}),\nn\\
\phi^{\prime 2} &=& \fft{c^2}{r^4f^2\rho^2} \left( \fft{f-\omega^2\rho^2}{r^4f\rho^2-c^2}\right) (r^4+R^4fh^{-1}r^{\prime 2}).
\eea
In order to have a well-defined string and avoid parts of it moving faster than the local speed of light, $-G$ must be positive definite all along the string. For nonzero $c$, this implies that $\phi^{\prime 2}>0$. Thus, the string spirals around the $\phi$ direction as it approaches the black hole, without reversing direction. 

In order for $-G>0$, any real roots of the numerator of $-G$ that lie along the string must also be roots of its denominator. Consider the function
\be\label{F}
F\equiv f-\omega^2\rho^2.
\ee
If there is a point $\rho=\rho_{crit}\le \bar\rho$ on the string such that $F(\rho_{crit})=0$, then this must specify the value of $c$ via
\be\label{c}
r_{crit}^4 f_{crit} \rho_{crit}^2-c^2=0,
\ee
where $r_{crit}\equiv r(\rho_{crit})$ and $f_{crit}\equiv f(\rho_{crit})$. Since $r_{crit}$ depends on $c$, the specified value of $c$ must be found numerically. On the other hand, if $\rho_{crit}>\bar\rho$, then $c$ is unspecified. In this case, given initial conditions with a nonvanishing value of $c$, the string will presumably evolve to the minimum-energy configuration with $c=0$.

\subsection{Static strings}

Let us first consider a static string with $\omega=0$. If the endpoint lies outside of the horizon ($\bar\rho>2M$) then, in order to have a well-defined string outside and on the horizon, we must have $c=0$. For the numerical integration, we take the boundary conditions to be at an intermediary point on the string $\rho=\rho_0$ and $r(\rho_0)=r_0$. Taking $\rho_0=2$ corresponds to a string which extends through the horizon, although the string endpoint is of course only sensitive to the portion of the string that lies outside of the horizon. In order to do the numerical integration, we have taken $\rho_0=2+10^{-8}$ for the region outside the horizon and then matched this with the region inside the horizon for $\rho_0=2-10^{-8}$. The numerical result for $\bar\rho=2.05$ and $R=M=b=1$ is given in the right plot of Figure \ref{fig2}. Note that this plot is no longer reliable as one gets too close to $\rho=0$. Of course, the string endpoint is only sensitive to the portion of the string that lies outside of the horizon.

\begin{figure}[h]
\centering
\includegraphics[width=160mm]{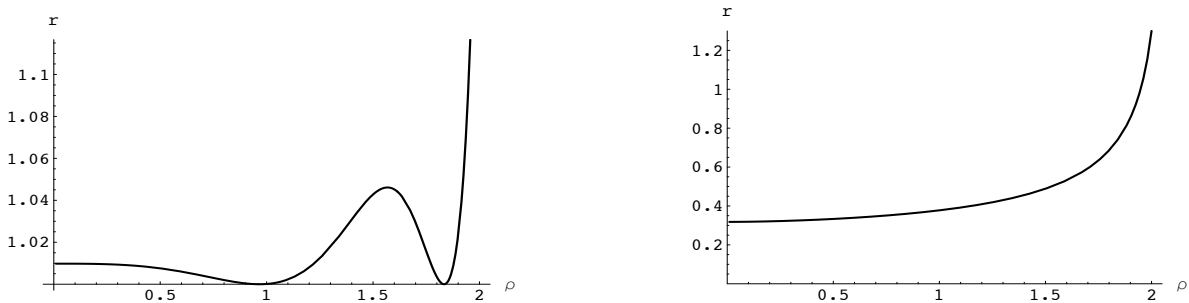}
\caption{The string extending through the horizon for $b=1$ (left) and $b=0$ (right).} 
\label{fig8}
\end{figure}

The left plot of Figure \ref{fig8} shows the portion of this string configuration within the horizon, from which we see that it undulates in the $r$ direction. This is a feature associated with the nonvanishing $b$ parameter of the AdS soliton. This is demonstrated by the right plot in Figure \ref{fig8}, for which we take $b=0$ and find the curve of the string to be monotonic in $r$. The field theory interpretation of these undulations could be that the interaction between the quark and the color charge of the black hole excites a collective gluonic resonance. Note however that this excitation is located within the event horizon and therefore cannot be detected by observers outside of the black hole.

\begin{figure}[h]
\centering
\includegraphics[width=60mm]{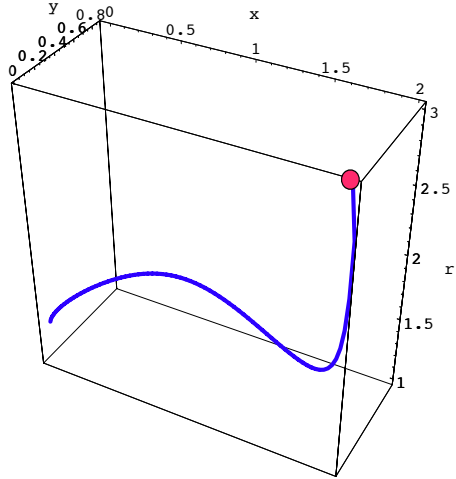}
\caption{A static string which lies entirely within the horizon and curves in the $\phi$ direction for $b=c=1$ and $\bar\rho=1.99$.} 
\label{fig9}
\end{figure}

If the string endpoint lies within the horizon ($\bar\rho<2M$), then this corresponds to a quark that is located within the event horizon of the black hole. Interestingly enough, in this case the point $\rho=\rho_c$ does not lie on the string. This means that the string is well-defined for arbitrary values of $c$, which corresponds to a string curving into the $\phi$ direction as it approaches the singularity. This is shown in Figure \ref{fig9} for the parameter values of $b=c=1$ for a string with an endpoint that lies just within the horizon at $\bar\rho=1.99$ and at $r_0=3$. Note that we are using the Cartesian coordinates $x=\rho \cos\phi$ and $y=\rho \sin\phi$.

\subsection{Steadily-rotating strings}

Now we will consider steadily-rotating strings. The constraints on $c$ can be determined from the roots of (\ref{F}), which is a cubic function of $\rho$. The discriminant is
\be
\Delta=4\omega^2 (1-27M^2\omega^2).
\ee
For $\omega> \omega_1\equiv 1/(\sqrt{27}M)$, $\Delta<0$ and there is one real root and two nonreal complex conjugate roots. Since $F(\rho\rightarrow -\infty)\rightarrow -\infty$ and $F(\rho\rightarrow 0^-)\rightarrow +\infty$, there is always a real negative root. Thus, there are no physical roots of $F$ for $\omega>\omega_1$. Then in order for the denominator of $-G$ to never vanish, we must have the lower bound $c^2\ge c_{min}^2$, where $c_{min}$ satisfies (\ref{c}). Note that $c_{min}\neq 0$ if the string endpoint is on or outside of the horizon, while $c_{min}=0$ if the endpoint lies within. In other words, for the case in which the string endpoint is outside of the horizon and moving at a high speed, there is a minimal amount with which the string must spiral around the $\phi$ direction.

For $\omega\le \omega_1$, (\ref{F}) has three real roots, which are all distinct for $\omega<\omega_1$ ($\Delta>0$) and two of which coincide for $\omega=\omega_1$ ($\Delta=0$). In particular, in addition to the real negative root, there are two real positive roots which lie outside of the  horizon for $\omega>0$. If one of them lies within the region $2M<\rho_{crit}\le \bar \rho$, then this specifies the value of $c$ via (\ref{c}). From (\ref{F}), we see that this occurs for $\omega\le \omega_2\equiv\sqrt{1-2M/\bar\rho}$. On the other hand, for $\omega>\omega_2$, then we must have the lower bound $c^2>c_{min}^2$. Note that the threshold velocity vanishes for quarks at the horizon and approaches the speed of light for quarks far from the black hole. 

To summarize, for a string endpoint moving with angular velocity $\omega$ whose endpoint lies outside of the horizon, the value of $c$ is specified for $\omega\le \mbox{Min}(\omega_1,\omega_2)$ while we have the lower bound $c^2>c_{min}^2$ for $\omega> \mbox{Min}(\omega_1,\omega_2)$, where $\mbox{Min}(\omega_1,\omega_2)$ is the minimum quantity between $\omega_1$ and $\omega_2$. On the other hand, if the endpoint lies within the horizon, then the value of $c$ can be arbitrary.

\begin{figure}[h]
\centering
\includegraphics[width=70mm]{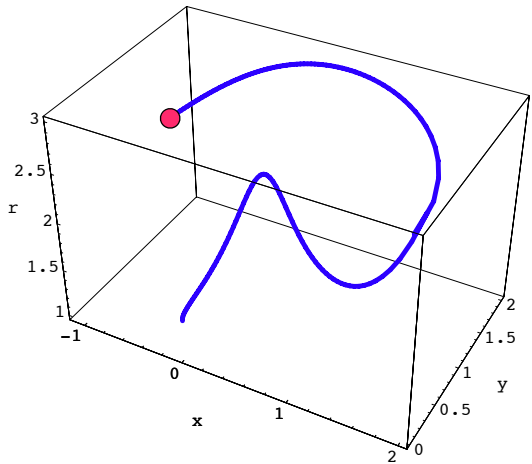}
\caption{A steadily-rotating, spiralling string for $\omega=0.5$, $b=c=1$ and $\bar\rho=2.02$.} 
\label{fig10}
\end{figure}

Figure \ref{fig10} shows a steadily-rotating, spiralling string whose endpoint lies outside of the horizon at $\bar\rho=2.02$ and is moving at $\omega=0.5$. We have taken the parameters $R=M=b=c=1$. Since $\omega>\mbox{Min}(\omega_1,\omega_2)\approx 0.1$, $c$ must satisfy $c^2>c_{min}^2$, which can be confirmed numerically. We have matched the solution outside of the horizon with $\rho_0=2.0005$ and $r_0=3$ to the solution within the horizon with $\rho_0=1.991$ and $r_0=3$. Spiralling strings near a black brane extended along the spacetime directions of the field theory have been considered in \cite{spiralling} and in AdS$_5$ in \cite{beaming1,beaming2,beaming3,beaming4}.

\begin{figure}[h]
\centering
\includegraphics[width=80mm]{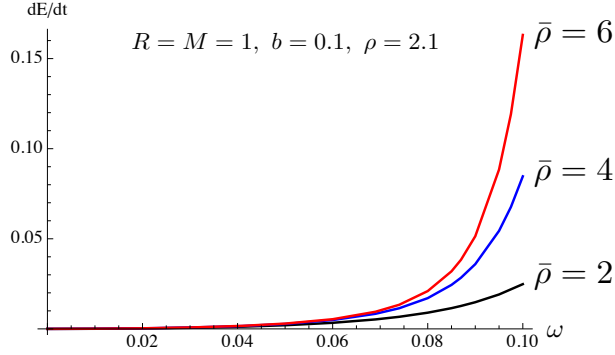}
\caption{Rate of energy loss versus angular velocity for various values of the mass parameter.} 
\label{fig11}
\end{figure}

As a quark circles around the black hole it loses energy, which is analogous to the energy loss of a quark moving in a strongly-coupled plasma \cite{herzog}. We have numerically computed the rate of energy loss as a function of the angular velocity $\omega$ for various values of the mass parameter $\bar\rho$, as shown in Figure \ref{fig11}. 
We used the shooting method in order to find the value of $c$ for a given $\omega$ that enforces the boundary condition $r(\bar\rho)=\bar r$. As can be seen, the rate of energy loss increases with angular velocity, as well as with the mass parameter. We have found that the results can be rather well fitted to the following formula:
\be
\fft{dE}{dt}\approx 0.008\omega \exp\left[ (20.65+16.45\bar\rho-0.25\bar\rho^2)\omega-0.89\bar\rho\right].
\ee
%

\section{Quark-Antiquark String Configurations}

\subsection{Static strings with endpoints separated in radial direction}

\begin{figure}[h]
\centering
\includegraphics[width=80mm]{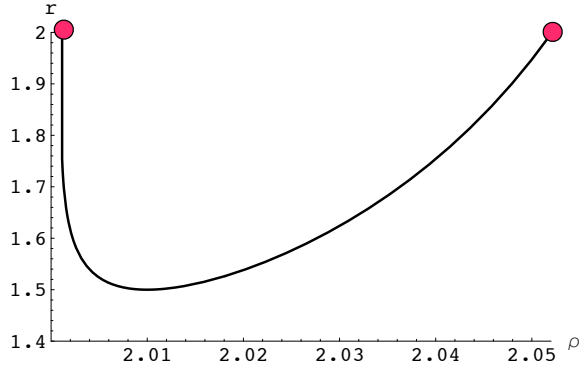}
\caption{A quark-antiquark string configuration with endpoints separated in the radial direction.} 
\label{fig1}
\end{figure}

A static quark-antiquark string with endpoints separated in the $\rho$ direction is described by the worldsheet embedding (\ref{embedding1}) with $\omega=0$, except now with the boundary conditions 
\be
0\le \tau\le T,\qquad \bar\rho_1\le\sigma\le\bar\rho_2,\qquad r(\bar\rho_1)=r(\bar\rho_2)=\bar r.
\ee
We will not consider light-heavy quark-antiquark pairs here, for which $r(\bar\rho_1)\ne r(\bar\rho_2)$. At the turning point, the equation of motion gives
\be\label{turning-point}
\partial_{\rho}^2r(\rho_0)=\fft{2r_0^3 h_0}{R^4 f_0},
\ee
where $f_0\equiv f(r_0)$ and $h_0\equiv h(r_0)$. For $r_0>b$, this is positive outside of the black hole which implies that outside of the black hole there are strings which descend below the probe brane and then turn back up. A numerical plot of a quark-antiquark string configuration is shown in Figure \ref{fig1}. The value of the turning point $r_0$ is chosen such that the string endpoint separation at $\bar r=2$ is kept fixed at approximately $0.05$, and we have taken the $\rho$ turning point to be at $\rho_0=2.01$. It can be seen that the midsection of the string is pulled towards the horizon, which is associated with a modification of the quark-antiquark potential due to the presence of the black hole.

\subsection{Static strings with endpoints separated in tangential direction}

A quark-antiquark pair separated along the $\phi$ direction can be described by the worldsheet embedding
\be
t=\tau,\qquad \rho=\rho(\sigma),\qquad \theta=\fft{\pi}{2},\qquad \phi=\sigma,\qquad r=r(\sigma),
\ee
with the boundary conditions
\be
0\le \tau\le T,\quad -\bar\phi_1\le\sigma\le \bar\phi_2,\quad r(\bar\phi_1)=r(\bar\phi_2) =\bar r,\quad \rho(\bar\phi_1)=\rho(\bar\phi_2)=\bar\rho.
\ee
Then the action becomes
\be
S=-\fft{T}{2\pi} \int d\phi \sqrt{\rho^2 \fft{r^4}{R^4} f+f h^{-1} r^{\prime 2}+\fft{r^4}{R^4} \rho^{\prime 2}},
\ee
where now $^{\prime}=\partial_{\phi}$. The equations of motion are
\bea\label{eom}
R^4 \fft{d}{d\phi}\left[ Af h^{-1}r^{\prime}\right] &=& 2r^3A\left( \rho^2f+ \rho^{\prime 2}
-\fft{5R^4 b^5 f}{2h^2 r^9} r^{\prime 2}\right),\nn\\
\fft{d}{d\phi} \left[ Ar^4 \rho^{\prime}\right] &=& A \left( r^4(\rho-M)+\fft{R^4 M}{h\rho^2} r^{\prime 2}\right),
\eea
where
\be
A^{-1}=\sqrt{\rho^2 r^4 f+R^4 f h^{-1} r^{\prime 2}+ r^4 \rho^{\prime 2}}.
\ee
Since the action does not depend on $\phi$ explicitly, the Beltrami identity yields
\be\label{beltrami}
A\rho^2 r^4 f=\rho_0 r_0^2 \sqrt{f_0},
\ee
where the turning point at the midsection of the string is taken to be at $\phi=0$, $r=r_0$ and $\rho=\rho_0$. This equation can be used in place of one of the equations of motion but it should still be verified that the resulting solution solves both equations of motion.

\begin{figure}[h]
\centering
\includegraphics[width=138mm]{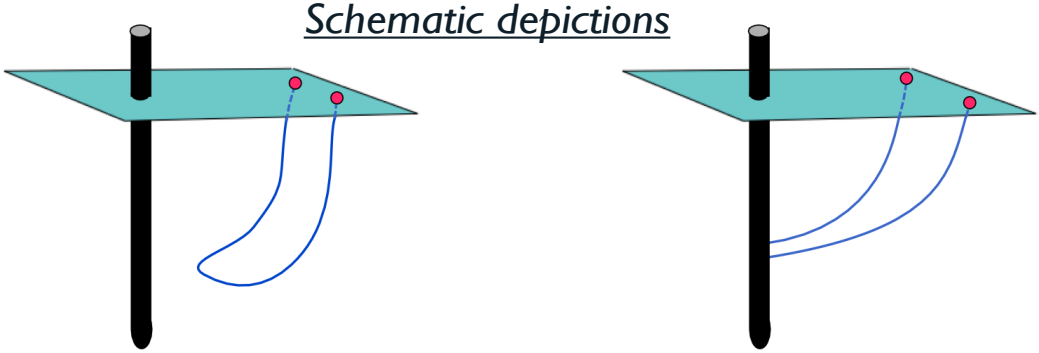}
\caption{Qualitative depictions of quark-antiquark string configurations whose endpoints are separated along the tangential direction for separation distance less than (left picture) and greater than (right picture) the screening length.} 
\label{fig3a}
\end{figure}

\begin{figure}[h]
\centering
\includegraphics[width=165mm]{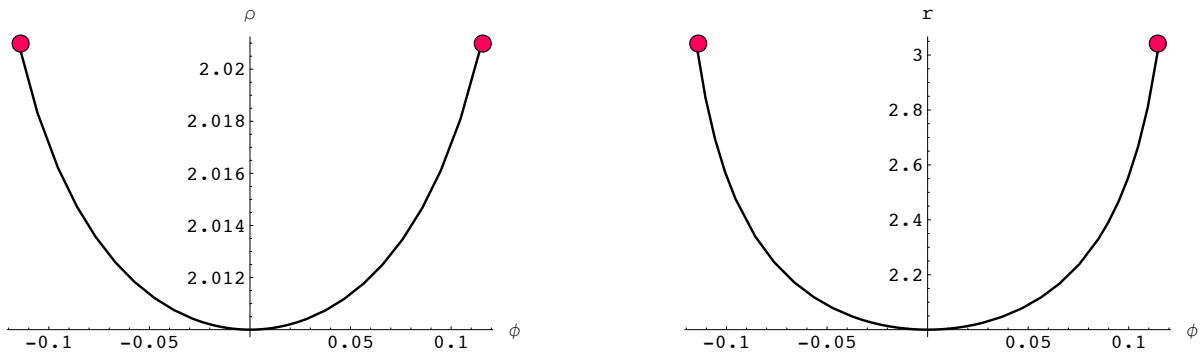}
\caption{The $r$ and $\rho$ coordinates as a function of angle $\phi$ for a quark-antiquark string configuration whose endpoints are separated along the tangential direction.} 
\label{fig3b}
\end{figure}

For the purpose of numerical integration, we take the boundary conditions to be at the midsection:
\be\label{mid-boundary}
r(0)=r_0,\qquad r^{\prime}(0)=0,\qquad \rho(0)=\rho_0,\qquad \rho^{\prime}(0)=0.
\ee
At this turning point, the equations of motion give
\be
r^{\prime\prime}(0)=\fft{2r_0^3\rho_0^2 h_0}{R^4}>0,\qquad \rho^{\prime\prime}(0)=\rho_0-M,
\ee
where $h_0>0$ provided that $r_0>b$. The first relation implies that the midsection of the string dips down. The second relation tells us that the midsection of the string dips towards the singularity at $\rho=0$ for $\rho_0>M$ and away from the singularity for $\rho_0<M$. A qualitative depiction of a string with one turning point is given by the left picture in Figure \ref{fig3a}, in which the bending of the string in the $\rho$ direction has been exaggerated. In reality, the bending of the string in the $\rho$ direction tends to occur at about two orders of magnitude less than the scale at which the string dips down in the $r$ direction. The two numerical plots in Figure \ref{fig3b} show the dependence of the $r$ and $\rho$ coordinates on $\phi$ for a string with $\rho_0=2.01$. 

If the distance between the string endpoints is increased, then at some point the energetically favorable configuration is that of two disconnected strings, which is schematically depicted in the right picture in Figure \ref{fig3a}. This is analogous to the screening length exhibited by a quark-antiquark pair in a strongly-coupled plasma \cite{screening1,screening2}, except in this case the quark-antiquark interaction is screened by the black hole.

There can be additional points along the string which are turning points in either the $r$ or $\rho$ directions. For instance, substituting (\ref{beltrami}) into the first equation in (\ref{eom}) for $r^{\prime}=0$ yields
\be
r^{\prime\prime}=\fft{2\rho^4 r^7 fh}{R^4\rho_0^2 r_0^4 f_0}.
\ee
This implies that the string dips down (up) in the vicinity of this $r$ turning point if it is located on the same (opposite) side of the horizon as the midpoint of the string. On the other hand, substituting (\ref{beltrami}) into the second equation in (\ref{eom}) for $\rho^{\prime}=0$ gives
\be\label{rho-eqn}
\rho^{\prime\prime}=\left[ 1+M\left( \fft{\rho r^4}{\rho_0^2r_0^4f_0}\right)\right] f\rho ,
\ee
thereby determining the regions in the $r-\rho$ plane for which the string curves away or towards $\rho=0$ in the vicinity of this $\rho$ turning point. 

\begin{figure}[h]
\centering
\includegraphics[width=165mm]{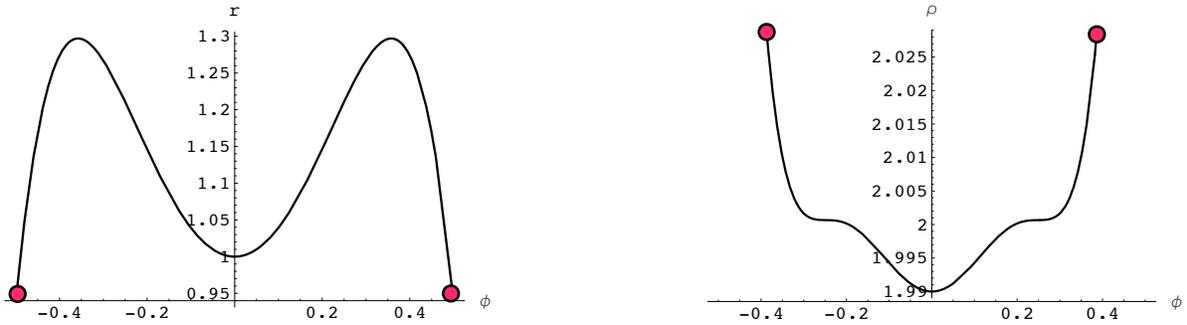}
\caption{The coordinates $r$ (left) and $\rho$ (right) for a quark-antiquark string configuration whose endpoints are outside the horizon and whose midsection lies within the horizon.} 
\label{fig10new}
\end{figure}

The behavior of the string is rather sensitive to its location within the $r-\rho$ plane. For instance, by changing the midsection turning point to be at $r_0=1$ and $\rho_0=1.99$, we find that there is a string configuration whose endpoints lie outside of the horizon and whose midsection lies inside of the horizon, as shown in the right plot of Figure \ref{fig10new}. The string exhibits saddlepoints on either side of its midsection where it then dips into the horizon. These saddlepoints occur since $\rho^{\prime\prime}=0$ at the horizon, as indicated by (\ref{rho-eqn}). From the left plot, of Figure \ref{fig10new} we see that the string lies above the endpoints in the $r$ direction, for which the two peaks occur outside of the horizon. Since the midsection of the string dips behind the horizon, the corresponding quark and antiquark is in some sense in an intermediate state between a bound quark-antiquark pair and that of free quarks. The peaks in the string on either side of its midsection could correspond to excitations of the quark potential due to the presence of the black hole, which serve to keep the quark-antiquark pair together.

\subsection{Steadily-rotating strings}

We will consider a quark-antiquark pair separated and steadily rotating in the $\phi$ direction. This can be described by the worldsheet embedding
\be
t=\tau,\qquad \rho=\rho(\sigma),\qquad \theta=\fft{\pi}{2},\qquad \phi=\sigma+\omega\tau,\qquad r=r(\sigma).
\ee
Then the action becomes
\be
S=-\fft{T}{2\pi} \int d\phi \sqrt{\rho^2 \fft{r^4}{R^4} f+\left(f-\omega^2\rho^2\right) \left(h^{-1} r^{\prime 2}+\fft{r^4}{R^4}f^{-1} \rho^{\prime 2}\right)},
\ee
where $^{\prime}=\partial_{\phi}$, as in the previous section. The equations of motion are
\bea\label{eom2}
R^4 \fft{d}{d\phi}\left[ B\left( f-\omega^2\rho^2\right) r^{\prime}\right] &=& 2r^3B\left[ \rho^2f+\left( 1-f^{-1}\omega^2\rho^2\right) \rho^{\prime 2}\right],\\
\fft{d}{d\phi} \left[ Br^4 \left( 1-f^{-1}\omega^2\rho^2\right) \rho^{\prime}\right]
&=& B \Big[ r^4(\rho-M)+\left( \fft{M}{\rho^2}-\omega^2\rho\right) R^4 r^{\prime 2}
+ f^{-2} \omega^2r^4 (3M-\rho)\rho^{\prime 2}\Big],\nn
\eea
where
\be
B^{-1}=\sqrt{\rho^2 r^4 f+\left( f-\omega^2\rho^2\right)\left(R^4 r^{\prime 2}+ r^4f^{-1} \rho^{\prime 2}\right)}.
\ee
Since the action does not depend on $\phi$ explicitly,
\be\label{beltrami2}
B\rho^2 r^4 f=\rho_0 r_0^2 \sqrt{f_0}.
\ee
We take the boundary conditions at $\phi=0$ to be given by (\ref{mid-boundary}). At the midsection turning point, the equations of motion give
\be
r^{\prime\prime}(0)=\fft{2r_0^3\rho_0^2}{R^4}\left( 1-f_0^{-1}\omega^2\rho_0^2\right)^{-1},\qquad \rho^{\prime\prime}(0)=(\rho_0-M)\left( 1-f_0^{-1} \omega^2\rho_0^2\right)^{-1}.
\ee
These relations tell us that the midsection of a string that is steadily moving at a high enough angular velocity and is located close enough to the horizon will curve in the opposite direction compared to that of a static string. In particular, the midsection of a string  located outside of the horizon and moving with an angular velocity of $\omega<\omega_{crit}$ ($\omega>\omega_{crit}$) bends towards (away from) $r=0$ and $\rho=0$, where $\omega_{crit}=\sqrt{f_0}/\rho_0$. 

As with static strings, there can be additional points which are turning points in either the $r$ or $\rho$ directions. To find a turning point in the $r$ direction, we substitute (\ref{beltrami2}) into the first equation in (\ref{eom2}) for $r^{\prime}=0$. This gives
\be
r^{\prime\prime}=\fft{2\rho^4 r^7 f}{R^4\rho_0^2 r_0^4 f_0}\left( 1-f^{-1}\omega^2\rho^2\right)^{-1}.
\ee
For a turning point in the $\rho$ direction, we substitute (\ref{beltrami2}) into the second equation in (\ref{eom2}) for $\rho^{\prime}=0$ to get
\be
\rho^{\prime\prime}=\left[ \rho-M+\left( M-\omega^2\rho^3\right) \left( \fft{\rho^2 r^4 f}{\rho_0^2 r_0^4 f_0}-1\right) \left( 1-f^{-1}\omega^2\rho^2\right)^{-1}\right] \left( 1-f^{-1}\omega^2\rho^2\right)^{-1}.
\ee
As with the midsection turning point, we see that the string bends in opposite directions near these points depending on whether $\omega<\sqrt{f}/\rho$ or $\omega>\sqrt{f}/\rho$ at the relevant turning point.

\section{Future Directions}

We have demonstrated that the AdS/CFT correspondence may be used to probe regions closer to the black hole singularity than expected from the four-dimensional gravity perspective. We show that quarks that interact strongly with a color-charged black hole share a number of the same features as quarks in a strongly-coupled plasma, such as quark energy loss and a quark-antiquark screening length.

There are a number of potentially useful and interesting future directions. Firstly, the stability of the Schwarzschild-black string AdS soliton still needs to be investigated. The AdS soliton has been shown to be perturbatively stable, its mass is lower than that of AdS itself and, moreover, there is evidence that it is the lowest energy state for particular boundary conditions \cite{lowest-mass1,lowest-mass2}. However, by no means does this indicate that the 
Schwarzschild-black string AdS soliton itself is stable.

It would be nice to extend our considerations to include the behavior of open strings on the generalizations of the Schwarzschild-black string AdS soliton that are discussed in the appendix, such as the one that describes the a strongly-coupled field theory near a rotating black hole. It would also be useful to consider color-charged black holes in a strongly-coupled plasma. A black hole in a strongly-coupled plasma has been conjectured to be dual to a ``black droplet'' or a ``black funnel'', depending on whether the plasma interacts weakly or strongly with the black hole, respectively \cite{hubeny,hubeny2,hubeny3}. Examples of these types of gravity backgrounds have been constructed in \cite{hubeny2,caldarelli}. 

Lastly, the AdS/CFT correspondence could enable us to probe the final stages of black hole evaporation to a greater extent than has been done with the semi-classical approach in four dimensions.

\appendix

\section{Various generalizations of the Schwarzschild-black string AdS soliton}

\subsection{Kerr-black string AdS soliton}

A rotating generalization of the Schwarzschild-black string AdS soliton consists of taking $ds_4^2$ to be the Kerr metric, which can be written as
\be
ds_4^2 = -\fft{\Lambda -a^2 \sin^2\theta}{\tilde\rho^2}\ dt^2+\fft{\tilde\rho^2}{\Lambda}\ d\rho^2+\tilde\rho^2 d\theta^2+\fft{\lambda \sin^2\theta}{\tilde\rho^2}\ d\phi^2+\fft{4Ma\rho \sin^2\theta}{\tilde\rho^2}\ dt d\phi,
\ee
where
\bea
\Lambda &=& \rho^2-2M\rho+a^2,\nn\\
\lambda &=& (\rho^2+a^2)^2-a^2\Lambda \sin^2\theta,\nn\\
\tilde\rho^2 &=& \rho^2+a^2\cos^2\theta.
\eea
In this case, the Kretschmann scalar is
\bea
K &=& \fft{12}{R^4}\left( 5+\fft{20b^{10}}{r^{10}}\right)\\
&+& \fft{12M^2R^4}{\tilde \rho^6 r^4} (\rho^2-a^2\cos^2\theta)\left[ 4\rho^4-28a^2\rho^2+a^4+a^2 \cos 2\theta (2\tilde\rho^2-30\rho^2+a^2)\right],\nn
\eea
from which we see that the geometry is smooth everywhere except at $\tilde\rho=0$, which corresponds to the usual ring singularity of the Kerr black hole.

\subsection{Schwarzschild-de Sitter-black string AdS soliton}

A solution whose constant-radius slices contains the four-dimensional Schwarzschild-de Sitter metric is given by
\be
ds_6^2 = \tilde h\ dy^2+\fft{r^2}{L^2} ds_4^2+\tilde h^{-1} dr^2,\nn\\
\ee
where
\be
\tilde h=1+\fft{r^2}{R^2}-\fft{b^3}{r^3},
\ee
where the de Sitter black hole metric can be written as (\ref{metric4}) with $f$ now given by
\be
f=1-\fft{2M}{\rho}-\fft{\rho^2}{L^2}.
\ee
The Kretschmann scalar is given by
\be
K=\fft{60}{R^4}+\fft{240b^6}{r^{10}}+\fft{48L^4M^2}{r^4\rho^6}.
\ee
 The solution is free of conical singularities if the coordinate $y$ has a periodicity of 
 \be
 \Delta y=\fft{4\pi R^2 c^4}{2c^5+3b^3R^2},
 \ee
 where $c$ is the largest root of the function $\tilde h$.

\subsection{In six-dimensional gauged supergravity}

The six-dimensional gauged supergravity constructed in \cite{6dsugra} has a solution which describes a black string AdS soliton supported by a single component of the $SU(2)$ Yang-Mills fields. This can be obtained from an analytical continuation of the AdS black hole solution \cite{6dbh} for the case in which the foliating surfaces of the transverse space is toroidal. The relevant AdS black hole solution can be written as
\bea
ds_6^2 &=& -H^{-3/2} h\ dt^2+H^{1/2} \left( h^{-1} dr^2+r^2 dx_i^2\right),\nn\\
\phi &=& \fft{1}{\sqrt{2}} \log H,\qquad A_{(1)}=\sqrt{2} \fft{(1-H^{-1})}{\sinh\beta}\ dt,\nn\\
h &=& -\fft{\mu}{r^3}+\fft29 g^2r^2 H^2,\qquad H=1+\fft{\mu \sinh^2\beta}{r^3},
\eea
where $i=1,\dots 4$. An AdS soliton can be obtained by taking the following analytical continuation of this solution:
\be
t\rightarrow iy,\qquad x_1\rightarrow it,\qquad \beta\rightarrow i\beta. 
\ee
This yields
\bea
ds_6^2 &=& \tilde H^{-3/2} \tilde h\ dy^2+\tilde H^{1/2} \left( \tilde h^{-1} dr^2+r^2 ds_4^2\right),\nn\\
\phi &=& \fft{1}{\sqrt{2}} \log \tilde H,\qquad A_{(1)}=\sqrt{2} \fft{(1-\tilde H^{-1})}{\sinh\beta}\ dy,\nn\\
\tilde h &=& -\fft{\mu}{r^3}+\fft29 g^2r^2 \tilde H^2,\qquad \tilde H=1-\fft{\mu \sinh^2\beta}{r^3},
\eea
where we can now replace $ds_4^2$ with a black hole metric in order to obtain a generalization of the Schwarzschild soliton string. The Kretschmann scalar is of the form
\be
K=\fft{P(r,\rho)}{\tilde H^5 r^{24}\rho^6},
\ee
where $P(r,\rho)$ is a polynomial of $r$ and $\rho$. Other than the singularity at $\rho=0$, the geometry is smooth for $r>c>(\mu\sinh^2\beta)^{1/3}$, where $c$ is the largest root of $\tilde h$. Also, the solution is free of conical singularities if the coordinate $y$ has a periodicity of $\Delta y=4\pi \tilde H(c)/\tilde h^{\prime}(c)$.

\section*{Acknowledgments}

We are grateful to Philip Argyres for useful discussions. This work is supported in part by NSF grant PHY-0969482 and a PSC-CUNY Award jointly funded by The Professional Staff Congress and The City University of New York.


\end{document}